\documentclass[aps,prc,reprint,nofootinbib]{revtex4-1}
\usepackage{graphicx}   %       for graphics
\usepackage{latexsym}   %       for special symbols
\usepackage{enumerate}
\usepackage{amsmath,amsfonts,amsthm} % Math packages
\usepackage{subfigure}
\begin{document}

\title{The cumulants of the baryon number from central Au+Au collision\\ at $E_{lab}$= 1.23 GeV$/$nucleon reveal the nuclear mean-field potentials}

\author{Yunxiao Ye$^{1,2}$}
\author{Yongjia Wang$^{2}$}
\email{wangyongjia@zjhu.edu.cn}
\author{Jan Steinheimer$^{3}$}
\author{Yasushi Nara$^{3,4}$}
\author{Hao-jie Xu$^{2}$}
\author{Pengcheng Li$^{2,5}$}
\author{Dinghui Lu$^{1}$}
\author{Qingfeng Li$^{2,6}$}
\email{liqf@zjhu.edu.cn}
\author{Horst Stoecker$^{3,7,8}$}

\affiliation{$^1$Department of Physics, Zhejiang University, Hangzhou 310027, China}
\affiliation{$^2$School of Science, Huzhou University, Huzhou 313000, China}
\affiliation{$^3$Frankfurt Institute for Advanced Studies, Ruth-Moufang-Str. 1, D-60438 Frankfurt am Main, Germany}
\affiliation{$^4$Akita International University, Yuwa, Akita-city 010-1292, Japan}
\affiliation{$^5$School of Nuclear Science and Technology, Lanzhou University, Lanzhou 730000, China}
\affiliation{$^6$Institute of Modern Physics, Chinese Academy of Sciences, Lanzhou 730000, China}
\affiliation{$^7$Institut f\"ur Theoretische Physik, Goethe Universit\"at Frankfurt, Max-von-Laue-Strasse 1, D-60438 Frankfurt am Main, Germany}
\affiliation{$^8$GSI Helmholtzzentrum f\"ur Schwerionenforschung GmbH, Planckstr. 1, D-64291 Darmstadt, Germany}

\begin{abstract}
Fluctuations of the baryon number in relativistic heavy-ion collisions are a promising observable to explore the structure of the QCD phase diagram. The cumulant ratios in heavy ion collisions at intermediate energies ($\sqrt{s_{\textrm{NN}}} < 7$ GeV) have not been studied to date.
We investigate the effects of mean field potential and clustering on the cumulant ratios of baryon and proton number distributions in Au+Au collisions at beam energy of 1.23 GeV$/$nucleon as measured by the HADES Collaboration at GSI.
Ultrarelativistic Quantum Molecular Dynamics (UrQMD) and the JAM model are used to calculate the cumulants with different mean field potentials.
It is found that the cumulant ratios are strongly time dependent.
At the early stage, the effects of the potentials on the fluctuations of the particle multiplicity in momentum space are relatively weak.
The mean fields enhance the fluctuations during the expansion stage, especially for small rapidity acceptance windows. The enhancement of cumulant ratios for free protons is strongly suppressed as compared to that for all baryons.
The mean field potentials and the clustering play an important role for the measured cumulant ratios at intermediate energy.
\end{abstract}

\maketitle

%----------------------------------------------------------------------------------------
%	CHAPTER 1
%----------------------------------------------------------------------------------------

\section{Introduction}

The major motivation to study relativistic heavy-ion collisions (HICs) is to explore the QCD phase diagram and to reveal the properties of the dense matter formed, e.g. of the quark-gluon plasma (QGP).
It is known from Lattice QCD - at zero baryon chemical potential ($\mu_B$) - that the transition from hadronic matter to a QGP is a smooth crossover.
Theoretical studies suggest that a first-order phase transition may exist at large baryon chemical potentials $\mu_B$, with a QCD critical point (the end point of the first-order phase boundary) at a certain temperature $T_C$ and $\mu_{C}$ \cite{signatures-CP,QCD phase diagram and the critical point,Critical point of QCD-lattice}.
So far, the existence of the conjectured critical point is an important open issue \cite{sign problem}. QCD matter off the ground state, can be created at different $T$ and $\mu_B$ by varying the colliding systems size, beam energy and impact parameters, to search for the critical endpoint.
This is one of the prime goals of the Beam-Energy Scan (BES) program at Relativistic Heavy Ion Collider (RHIC) \cite{Aggarwal:2010cw,Kumar:2013cqa,Adamczyk:2017iwn,BESII,Yang:2017llt}, the NA61 experiment at the CERN-SPS \cite{NA61SHINE}, the HADES and CBM experiments at GSI and FAIR \cite{FAIR}, as well as dedicated future programs at NICA \cite{NICA} and J-PARC \cite{JPARC-HI}.

To determine the critical point from HICs, the fluctuations of conserved charges, which are sensitive to the correlation length $\xi$ in (QCD-)matter, have been conjectured as a promising observable~\cite{Event-by-event fluctuations,sensitive observable}. The divergence of $\xi$ results in critical phenomena near the critical point, see e.g.~\cite{sign of kurtosis,sign of third moments, lattice predictions of sign,Bzdak2012,Konchakovski2009,Bzdak2016,Kitazawa2016,Mukherjee2015,Vovchenko2015,Cheng2009,Morita2015,Borsnyi2013} .

The STAR Collaboration has measured the fluctuations of net-proton, net-charge and net-kaon number in Au+Au collisions from $\sqrt{s_{NN}}$ = 7.7 to 200 GeV.
A flat beam-energy dependence of net-charge and net-kaon number fluctuations was observed, while preliminary data on the kurtosis ($\kappa\sigma^2$) of net-protons show a non-monotonic behavior as a function of beam-energy in the most central Au+Au collisions \cite{STAR-proton,STAR-charge,PHENIX-experiment-net-charge,Xu2014}.
This interesting non-monotonic behavior is a strong motivation for the BES-II program at RHIC: The STAR collaboration has proposed to measure cumulants in Au+Au collision from $\sqrt{s_{NN}}$ =7.7 to 19.6 GeV. Fixed target experiments from $\sqrt{s_{NN}}=2.7$ to 4.9 GeV have been proposed at the Compressed Baryonic Matter (CBM) detector at the future Facility for Antiproton and Ion Research (FAIR)\cite{GSI} adjacent to GSI.

At intermediate energies, from 0.1 to 2 GeV$/$nucleon, at the present Schwer-Ionen Synchrotron (SIS) accelerator at GSI, the higher moments of the proton number distribution in Au+Au collisions at beam energy of 1.23 GeV$/$nucleon have been measured by the HADES collaboration.
The interpretation of the data and a comparison to data at higher beam energies is presently discussed.
At SIS18 energies, nuclear matter with densities of twice to three times saturation density is created and a large fraction of the emitted protons and neutrons at midrapidity is bound in fragments.
Both the collective flow ($v_1$ and $v_2$) and the baryon stopping reach their maximum here.
Thus, the higher moments of the proton number distribution at SIS energies are more complicated to evaluate.
Higher moments of the net-proton number distribution are also influenced by other effects, like system volume fluctuations \cite{skokov2013,haojie-2016}, efficiency corrections \cite{bzdak2015,kitazawa2016}, baryon clustering \cite{Shuryak:2018lgd}, global charge conservation \cite{Sakaida:2014pya}, etc.
The ultra-relativistic quantum molecular dynamics (UrQMD) model has been used to study the higher order cumulants of net-protons in Au+Au collisions at a beam energy of $E_{lab}$=1.23 GeV$/$nucleon.
The nuclear interactions have shown sizable effects on the cumulant ratios \cite{jan2018}.\\

In the present work, the multiplicity distributions of baryons and protons at central Au+Au collisions at beam energy $E_{lab}$=1.23 GeV/nucleon are calculated with the UrQMD model and different mean-field potentials and nuclear clustering effects are studied.
The difference of the cumulants calculated for all baryons and free baryons (where baryons inside clusters are subtracted) is found to be large.

%----------------------------------------------------------------------------------------
%	CHAPTER 2
%----------------------------------------------------------------------------------------

\section{The UrQMD model}

The UrQMD  model is a microscopic many-body transport approach in which each hadron is represented by a Gaussian wave packet in phase space. The
time evolution of the centroids ($\textbf{r}_i$ and $\textbf{p}_i$) of the Gaussians obey Hamilton's equations,
\begin{eqnarray}\label{eq1}
\dot{\textbf{r}}_{i}=\frac{\partial  \langle H  \rangle}{\partial\textbf{ p}_{i}},
\dot{\textbf{p}}_{i}=-\frac{\partial  \langle H \rangle}{\partial \textbf{r}_{i}}.
\end{eqnarray}
Here {\it $\langle H \rangle$} is the total Hamiltonian function of the system, it consists of the kinetic energy of the particles and the effective interaction potential energy. The importance of the mean field potential for describing HICs has been extensively studied \cite{Stoecker:1986ci,Bertsch:1988ik,Aichelin:1987ti,Gale:1987zz,Bertsch:1988ik}. For studying HICs at SIS energies, the following density and momentum dependent potential has been widely used \cite{Aichelin:1991xy,Hartnack:1997ez,Li:2005gfa},
\begin{equation}\label{eq2}
U=\alpha (\frac{\rho}{\rho_0})+\beta (\frac{\rho}{\rho_0})^{\gamma} + t_{md} \ln^2[1+a_{md}(\textbf{p}_{i}-\textbf{p}_{j})^2]\frac{\rho}{\rho_0}.
\end{equation}
Here $\alpha$, $\beta$, $\gamma$, $t_{md}$, and $a_{md}$ are parameters which can be adjusted to yield a different nuclear incompressibility ($K_0$) for isospin symmetric nuclear matter. In this treatment, the gradient (see eqn.(\ref{eq1})) of the net-baryon density $\rho$, effectively introduces a finite-range interaction through the treatment of baryons as Gaussian wave packets. This finite-range interaction is attractive for dilute systems and repulsive for dense systems. Thus, it can be essential for the formation of correlations over long distances.

In order to study the influence of different nuclear mean field potentials on higher moments of the multiplicity distribution, the so-called soft and momentum dependent (SM), hard and momentum dependent (HM), as well as the hard and momentum independent (H) nuclear potentials are chosen. The set of parameters are displayed in Table I. Those parameter sets have been widely used in studying mean field potential effects in HICs at intermediate energies~\cite{MF-parameters}.

\begin{table}[htbp]
\centering
\caption{\label{tab:table1}
Parameter sets of the nuclear equation of state.}
\setlength{\tabcolsep}{1.4pt}
\begin{tabular}{|l|cccccc|}
\hline
EoS & $K_0$(MeV) & $\alpha$(MeV) & $\beta$(MeV) & $\gamma$ & $t_{md}$(MeV) & $a_{md}$($\frac{c^2}{GeV^2}$) \\ \hline
SM & 200 & -393 & 320 & 1.14 & 1.57 & 500 \\
H & 300 & -165 & 126 & 1.676 & - & - \\
HM & 380 & -138 & 60 & 2.084 & 1.57 & 500 \\ \hline
\end{tabular}
\end{table}

Besides the nuclear mean-field potential, a short range stochastic scattering between two particles is also necessary in the transport model to compensate the strong repulsive short-range component of the nuclear interaction. It is well known that the in-medium nucleon-nucleon elastic cross section should be smaller than the free one~\cite{Li:2014oda,OL13,Li:2018wpv}.
Thus, in the present version of the UrQMD model, a density and momentum dependent in-medium correction factor on the free elastic cross section is applied. Details about the in-medium nucleon-nucleon cross section can be found in Ref~\cite{wyj1417}. At SIS energies, a large fraction of protons (neutrons) is bound in light fragments, e.g., in central Au+Au collision at $E_{lab}$=1.2 GeV/nucleon, the percentage of protons bound into clusters is about 40\%~\cite{FOPI:2010aa}. Therefore, a proper treatment of the clustering process is necessary.
In this work, an isospin dependent minimum spanning tree method \cite{iso-MST algorithm} is used to recognize nuclear clusters at the end of the simulation. In this method, if the relative distance between two protons or two neutrons (neutron and protons) is smaller than 2.8 fm or 3.8 fm, and the relative momentum is smaller than 0.25 GeV/$c$, they are considered to belong to the same cluster.
It has been found that by properly adjusting these parameters, the fragment mass distribution in intermediate energy HICs can be reproduced~\cite{hk31,scpma58,pp2011,Peilert:1989kr}.

%----------------------------------------------------------------------------------------
%	CHAPTER 3
%----------------------------------------------------------------------------------------

\section{Fluctuations}

In the grand-canonical ensemble, fluctuations of conserved charges can be characterized by susceptibilities which are the derivatives of the partition function $\ln Z$ with respect to the corresponding chemical potential,
\begin{equation}
\chi^q_i = \frac{\partial[\ln Z (V,T,\mu_q)/VT^3]}{\partial[\mu_q/T]^i}.
\end{equation}
In the grand-canonical ensemble these are related to the cumulants of particle multiplicity distributions on an event-by-event basis:
\begin{align}
\begin{split}
&C_1=M=\langle N \rangle, \\
&C_2=\sigma^2=\langle (\delta N)^2\rangle, \\
&C_3=S\sigma^3=\langle (\delta N)^3\rangle.
%&C_4=\kappa \sigma^4=\langle (\delta N)^4\rangle-3\langle (\delta N)^2 \rangle^2.
\end{split}
\end{align}
Here $\delta N=N- \langle N \rangle$ with $N$ being the number of particles in a given acceptance window (e.g., the rapidity or the transverse momentum window) for a single event.
$M$ is the mean value, $\sigma$ is the standard deviation and $S$ is the skewness, which measures the degree of asymmetry of a distribution. Usually, the ratios of cumulants are constructed to cancel the unknown volume dependence and directly compared with theoretical calculations of susceptibilities,
\begin{align}
\begin{split}
&C_2/C_1=\sigma^2/M,\\
&C_3/C_1=S \sigma^3/M,\\
&C_3/C_2=S \sigma.
\end{split}
\end{align}
According to the Delta-theorem \cite{error estimation}, the statistical error of the cumulants and their ratios can be approximated as follows:
\begin{align}
\begin{split}
&error(C_r) \propto \sigma^r/ \sqrt{n},\\
&error(C_r/C_2) \propto \sigma^{(r-2)}/ \sqrt{n}.
\end{split}
\end{align}
Here $n$ is the total number of events.

%----------------------------------------------------------------------------------------
%	CHAPTER 4
%----------------------------------------------------------------------------------------

\section{Numerical results}

The degree of stopping reaches a maximum at SIS energies. Hence, we first investigate the influence of different mean-field potentials on the stopping~\cite{FOPI:2010aa,WR23,Yingyuan}. The degree of stopping can be measured by \textit{varxz}, the ratio of the width of the transverse (usually refers to the \textit{x}-direction) rapidity distribution over that of the longitudinal (the \textit{z}-direction) rapidity distributions, defined as\cite{FOPI:2010aa},
\begin{equation}
varxz=\frac{<y_{x}^2>}{<y_{z}^2>}. \label{eqvartl}
\end{equation}
Here
\begin{equation}
\left<y_{x,z}^2\right>=\frac{\sum(y^2_{x,z}N_{y_{x,z}})}{\sum
N_{y_{x,z}}}, \label{eqgm}
\end{equation}
where $\left<y_{x}^2\right>$ and $\left<y_{z}^2\right>$ are the
widths of the rapidity distributions of particles in the $x$ and
$z$ directions, respectively. $N_{y_{x}}$ and
$N_{y_{z}}$ denote the numbers of particles in each $y_x$
and $y_z$ bins. Thus, in the case of full stopping: $varxz$=1, while full transparency yields $varxz$=0.

\begin{figure}[htbp]
\centering
\includegraphics[angle=0,width=0.5\textwidth]{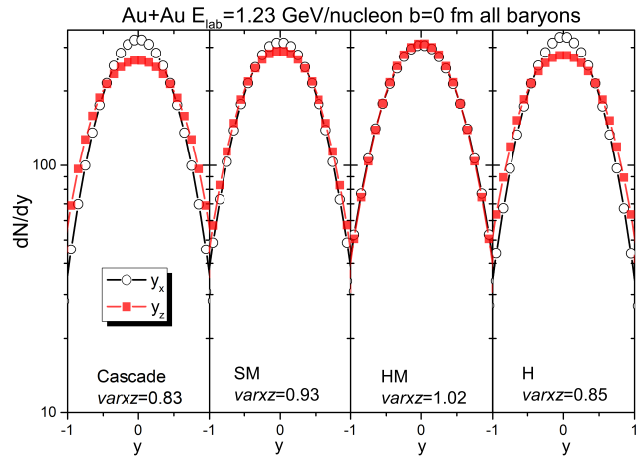}
\caption{\label{CS-fig1} Yield distributions of all baryons as functions of the reduced longitudinal ($y_z$/$y_{b}$, filled squares) and transverse ($y_x$/$y_{b}$, open circles) rapidities from central Au+Au collisions at beam energy of 1.23 GeV$/$nucleon. Calculations with the soft momentum-dependent (SM), the hard momentum-dependent (HM), and the hard without momentum-dependent (H) mean field potentials are compared to calculation without mean field potential (cascade mode). The corresponding values of $varxz$ are also shown. The error bars are smaller than the symbols size.}
\end{figure}

In this work, we focus on central Au+Au collisions at a beam energy of 1.23 GeV$/$nucleon, experimentally measured by the HADES collaboration at GSI.
For each of the different potentials presented, 2 million events were simulated.
The total propagation time was 100 fm$/$c, unless explicitly stated otherwise. Fig.\ref{CS-fig1} shows the yield distributions of all baryons for head-on (b=0 fm) Au+Au collisions. Clearly, the rapidity distributions are influenced by the mean field potential.
The degree of stopping is much larger for the HM than that in the cascade mode, due to the strong repulsive interaction at the high density phase, in the case of the HM. It is evident that the mean field potential also has a strong effect on the clustering. This will influence the measurable cumulant ratios, as we will see later.

%----------------------------------------------------------------------------------------

\begin{figure}[t]
\centering
\includegraphics[width=0.5\textwidth]{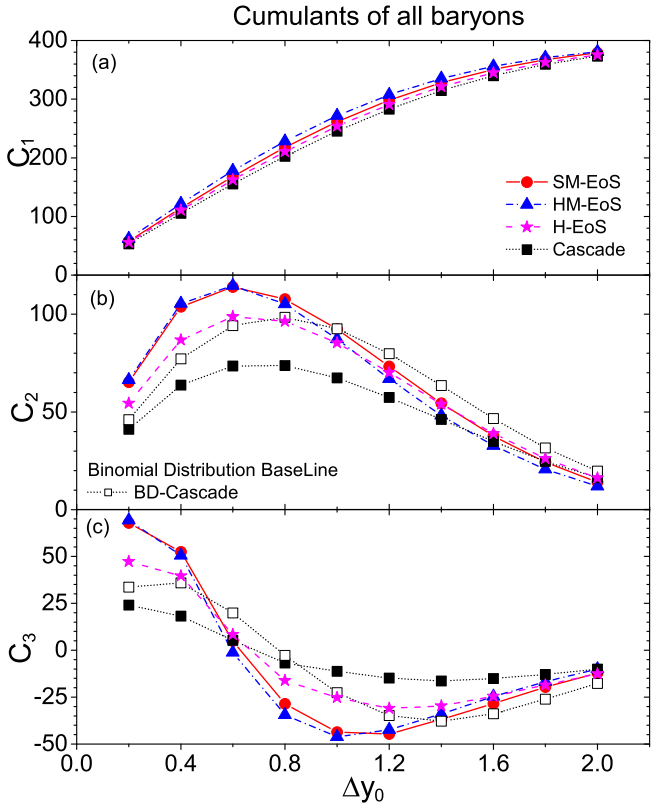}
\caption{\label{fig2} Rapidity dependence for the cumulants ($C_1$ - $C_3$) of all baryons produced in central Au+Au collisions at beam energy of 1.23 GeV/nucleon. Results have been calculated with SM, HM and H potentials as well as in the cascade mode (solid symbols). The results are compared to a baseline from the binomial distribution (open symbols). Error bars for all data shown, are smaller than the symbol sizes.}
\end{figure}

\begin{figure}[t]
\centering
\includegraphics[width=0.5\textwidth]{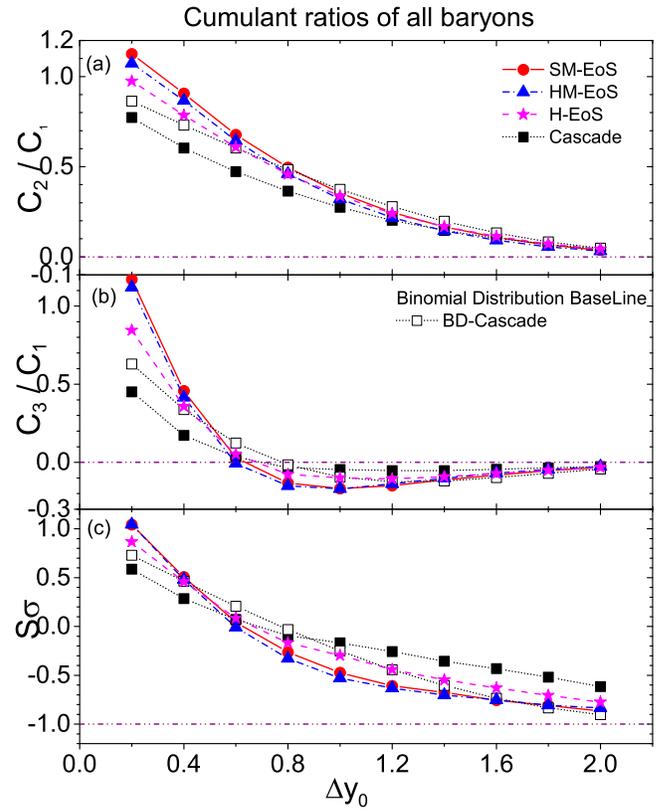}
\caption{\label{fig3}
Rapidity dependence for the cumulant ratios ($C_2$/$C_1$, $C_3$/$C_1$, $S\sigma$) of all baryons produced in central Au+Au collisions at beam energy of 1.23 GeV/nucleon. Results have been calculated with SM, HM and H potentials as well as in the cascade mode (solid symbols). The results are compared to a baseline from the binomial distribution (open symbols). Error bars for all data shown, are smaller than the symbol sizes.}
\end{figure}

\subsection{Final results on baryon number cumulants}

Figs. \ref{fig2} and \ref{fig3} display the cumulants and their ratios for all baryons
in different rapidity window $\Delta y_0$ (around mid-rapidity). Here
$y_0$ is the scaled rapidity divided by the beam rapidity in the c.m. frame of the collision:
$y_0=y/y_\text{b}$.

We also compare the cumulants from the transport simulations with a Binomial baseline. This baseline essentially assumes uncorrelated baryon emission, while enforcing global baryon number conservation.
$C_2$ and $C_3$ for the binomial distribution are obtained by
\begin{eqnarray}
C_2&=&Np(1-p) \\
C_3&=&Np(1-2p)(1-p),
\end{eqnarray}

with $N$=394 being the total baryon number and $p$=$C_1$/$N$ being the average fraction of baryons in the given acceptance. Here $C_1$ is taken from simulation in the cascade mode. We have found that the baseline of the binomial distribution will not change much if $C_1$ is taken from different scenarios. Hence only the baselines of the cascade mode are shown.\\

In general, the results obtained for different mean field potentials as well as those without mean field potential (cascade mode) have analog features (e.g., $C_2$ first increases with the rapidity window up to a maximum then decreases.
Both $C_2$/$C_1$ and $S\sigma$ show a monotonous decrease with the rapidity window, while $C_3$/$C_1$ first decreases with increasing acceptance, up to a minimum and then increases). On the other hand the magnitude of the change of the cumulants and their ratios varies drastically for the different potential implementations. While the momentum dependent potentials essentially give the same result, they also show the largest deviation from the binomial baseline. The momentum independent potentials give results which are closer to those of the cascade version of the model.
In particular, for small rapidity windows $\Delta y_0 \approx 0.4$, $C_2$/$C_1$, $C_3$/$C_1$, and $S\sigma$ calculated with mean field potentials (i.e., SM, HM, and H) are larger than the binomial baseline, while the ones calculated in the cascade mode are smaller than the binomial baseline.\\

\begin{figure}[t]
\includegraphics[width=0.5\textwidth]{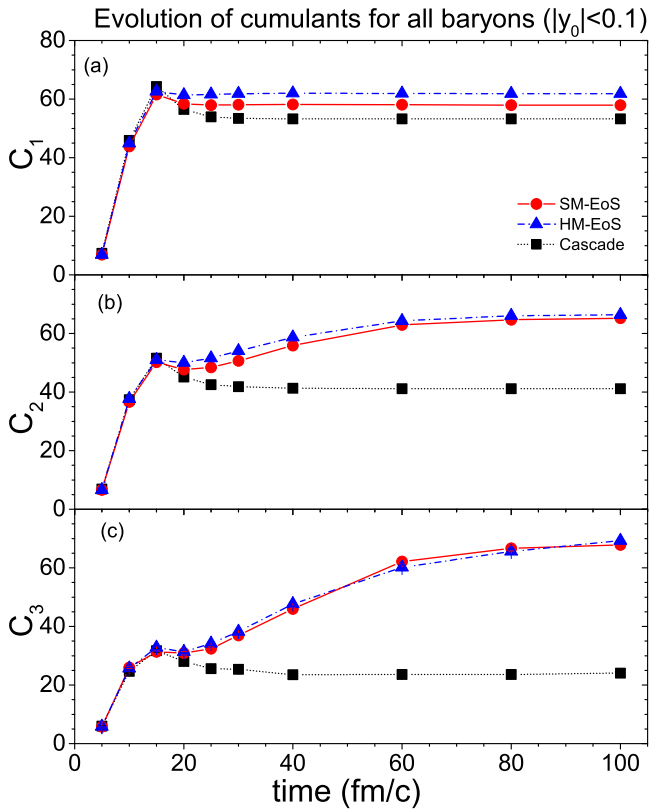}
\caption{\label{fig4} Time evolution for the cumulants ($C_1$, $C_2$ and $C_3$) of all baryons at mid-rapidity produced from central Au+Au collisions at beam energy of 1.23 GeV/nucleon. Results have been calculated with HM and SM potentials as well as in the cascade mode (solid symbols). Error bars are smaller than symbols size.}
\end{figure}

\begin{figure}[t]
\includegraphics[width=0.5\textwidth]{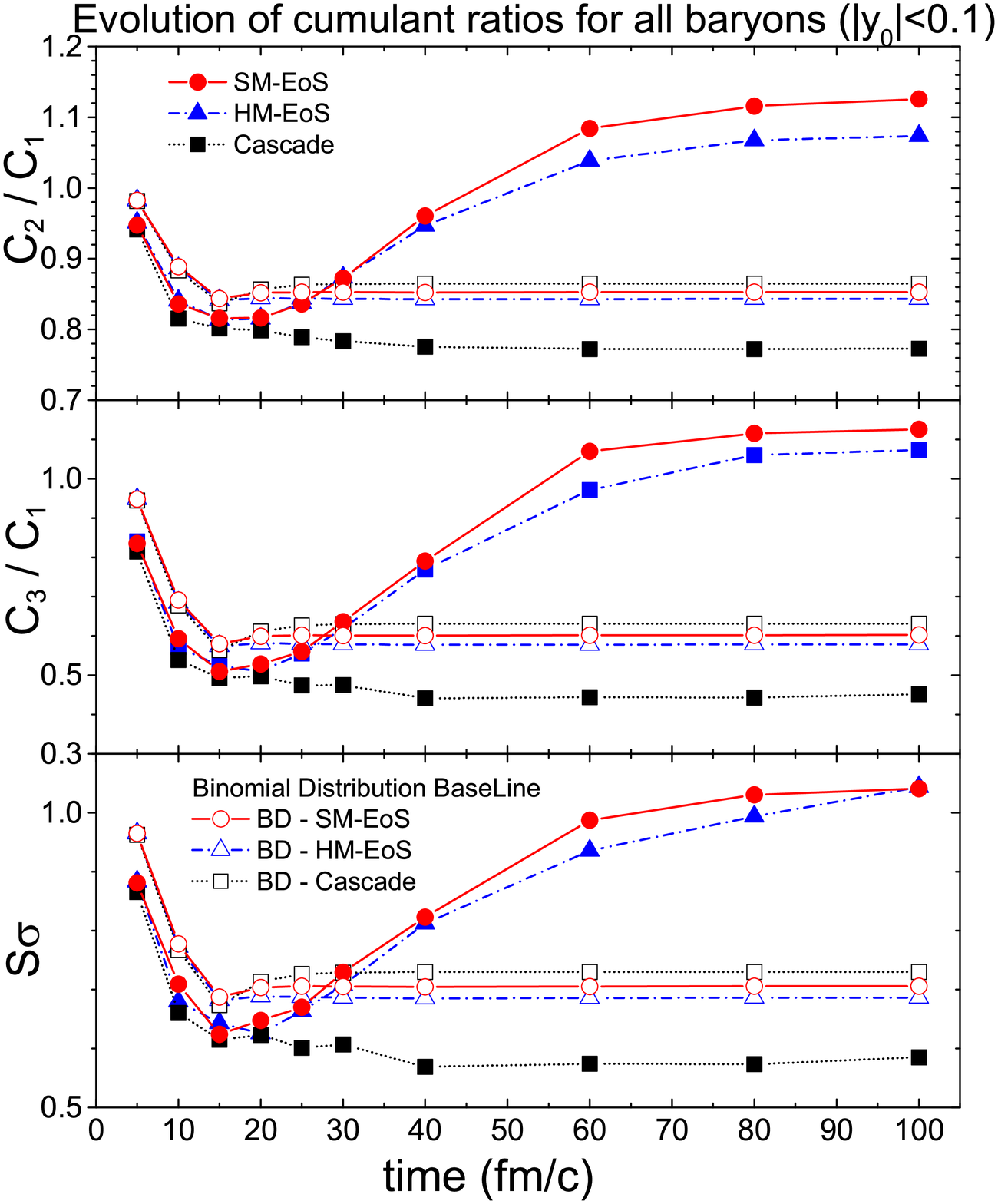}
\caption{\label{fig5} Time evolution for the cumulants ratios ($C_2$/$C_1$, $C_3$/$C_1$ and $S\sigma$) of all baryons at mid-rapidity produced from central Au+Au collisions at beam energy of 1.23 GeV/nucleon. Result have been calculated with HM and SM potentials as well as in the cascade mode (solid symbols). Error bars are smaller than symbols size. The results are compared to the baseline of the binomial distribution (open symbols).}
\end{figure}

The increased cumulant ratios in calculations with the mean field potentials indicates that the nuclear interaction enlarge the correlation (i.e., the fluctuation of $\delta N$) in each rapidity window.
For larger rapidity windows, the differences in the cumulant ratios among different calculations steadily decrease, and their values approach the limiting values obtained from the binomial distribution for $p=1$, i.e., $C_2$/$C_1$=$C_3$/$C_1$=0 and $S\sigma=-1$, due to the dominant contribution from baryon conservation. In addition, as $\Delta y_0$ becomes larger than the correlation length of the potential interaction, the effect will also decrease.

The fact that the cumulant ratios calculated with SM and HM are very close to each other even though the difference in the nuclear incompressibility $K_0$ is as large as 180 MeV, while the results obtained with H do not track closely with the results of HM illustrate, that the cumulant ratios are less sensitive to the incompressibility $K_0$ but more sensitive to the momentum-dependent component of the nuclear potential.

%-----------------------------------------------------------------------------------------

\subsection{Understanding the time Evolution of the Cumulants}

In order to better understand the mean field effects on the cumulant ratios, the results at various time points are displayed in Fig.~\ref{fig4} and Fig.~\ref{fig5} . The rapidity window $|y_{0}|$$\le$0.1 is chosen to weaken the influences of baryon number conservation. Comparing with the binomial baseline, the different deviations imply different correlations.
In the early stage ($t$$\le$15 fm$/$c), which corresponds to the compression period, the cumulant ratios obtained from HM, SM and cascade mode are very close to each other, and decrease with increasing time because of the increased baryon number in the mid-rapidity region.
Our previous work \cite{jan2018} found that the cumulant ratios in the coordinate space are significantly influenced by the mean field potential at an early stage. This makes sense, as the correlations are first space-like and need to be transformed to momentum-space correlations at a later time.
At $t$$\ge$15 fm$/$c, the compressed matter begins to expand, the magnitudes of the cumulant ratios obtained with the mean fields (HM and SM) increase with increasing time and saturate at a larger value, while these obtained with the cascade mode as well as the binomial distribution baseline almost remain constant. The enhanced magnitude of the cumulant ratios in the presence of mean field potential also can be observed in a larger rapidity acceptance window.
Since at late times most of the collision have ceased, the momentum of particles will not be modified too much in the absence of mean field potential, thus the cumulant ratios remain constant in the cascade mode.
In the presence of mean field potential (SM and HM), the momentum of particles could be influenced by surrounding particles through the nuclear interaction.
Fig~\ref{fig4} and \ref{fig5} show that the nuclear interaction will enhance the momentum correlation of nucleons in the freeze-out stage.
When the system becomes dilute at late times; sub-saturation
density are reached; the long range attractive interaction dominates, which
leads to a positive correlation between baryons,
thus it increases the $C_2/C_1$.

In addition, the results for SM increase faster than that for HM, because SM yields a stronger attractive potential at low densities.

\begin{figure}[t]
\centering
\includegraphics[angle=0,width=0.5\textwidth]{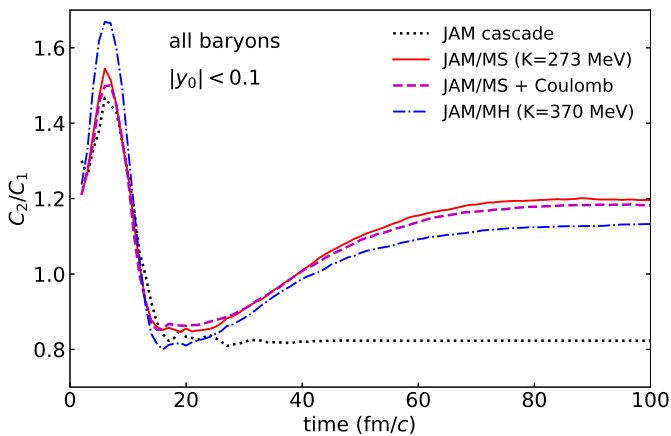}
\caption{\label{Jam} Time evolution for $C_2/C_1$ of all baryons from JAM. Calculations with momentum-dependent hard (MH) and soft (MS) potentials are compared to the result obtained with cascade mode.
}
\end{figure}

\subsection{Model Dependence}
The extracted values of the cumulants may depend on the parameters and details of the potential implementation.
To study the model dependence on the cumulants, we plot
in Fig.~\ref{Jam}, the $C_2/C_1$ of all baryons from the transport model JAM.
A detailed description of JAM can be found in Ref.~\cite{JAM}.
We use the parameter set for the potentials
which was found in Ref.~\cite{JAM2}.
Previously, JAM was applied for higher beam energies ($\sqrt{s_{NN}}=5$ GeV),
and it was found that nuclear potential effects are very small for the cumulants%
~\cite{He:2016uei} at these energies.
The implementation of hadronic mean-field in JAM is different from that in UrQMD;
JAM uses the same Skyrme-type density dependent potential as UrQMD,
but it uses the Lorentzian-type momentum dependent potential.
In addition, potentials in JAM are implemented as scalar, based on the simplified version of relativistic molecular dynamics (RQMD/S)~\cite{Maruyama:1996rn,Mancusi:2009zz}.

Regardless of the difference of the detailed implementations,
the results from the two models are almost entirely consistent with each other, i.e, after $t$$\ge$20 fm$/$c, $C_2/C_1$ increases with increasing time in the presence of mean field potential, while it remains constant in the calculation without mean field potential. The hard potential results in a smaller value of $C_2/C_1$ than the soft one. These similar results from two transport models manifest that mean field potential plays important role on $C_2/C_1$, while other physical ingredients of the transport models do not significantly affect $C_2/C_1$. Moreover, it can be seen that the effect of the Coulomb potential on $C_2/C_1$ of all baryons is very small.

\subsection{Effects of cluster formation}
%-------------------------------------------------------------------------------------

\begin{figure}[t]
\includegraphics[width=0.5\textwidth]{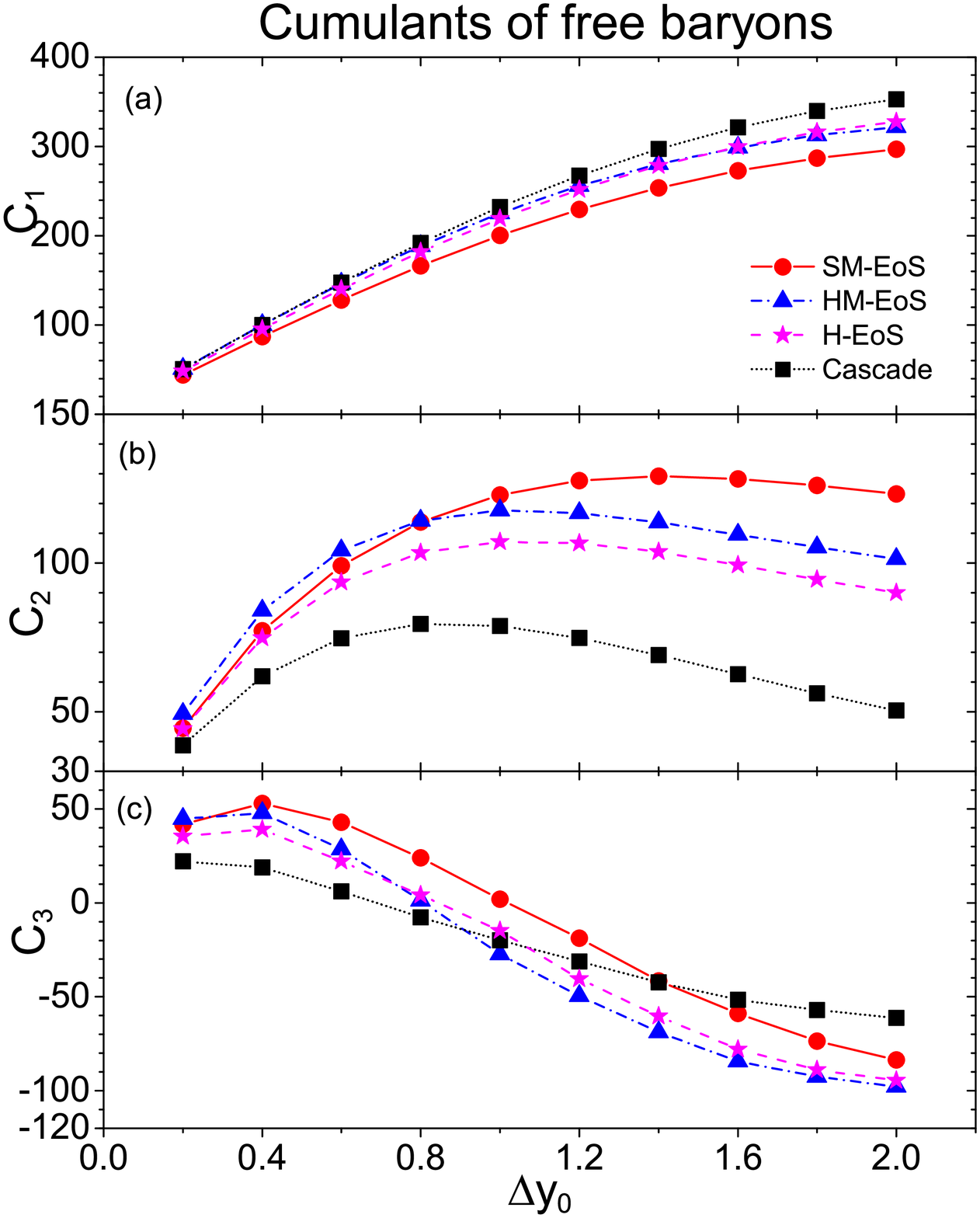}
\caption{\label{fig7} Rapidity dependence for the cumulants ($C_1$ - $C_3$) of free baryons, i.e. excluding all baryons that are in a cluster, produced in central Au+Au collisions at beam energy of 1.23 GeV/nucleon. Results have been calculated with SM, HM and H potentials as well as in the cascade mode (solid symbols). Error bars for all data shown, are smaller than the symbol sizes. }
\end{figure}

\begin{figure}[t]
\includegraphics[width=0.5\textwidth]{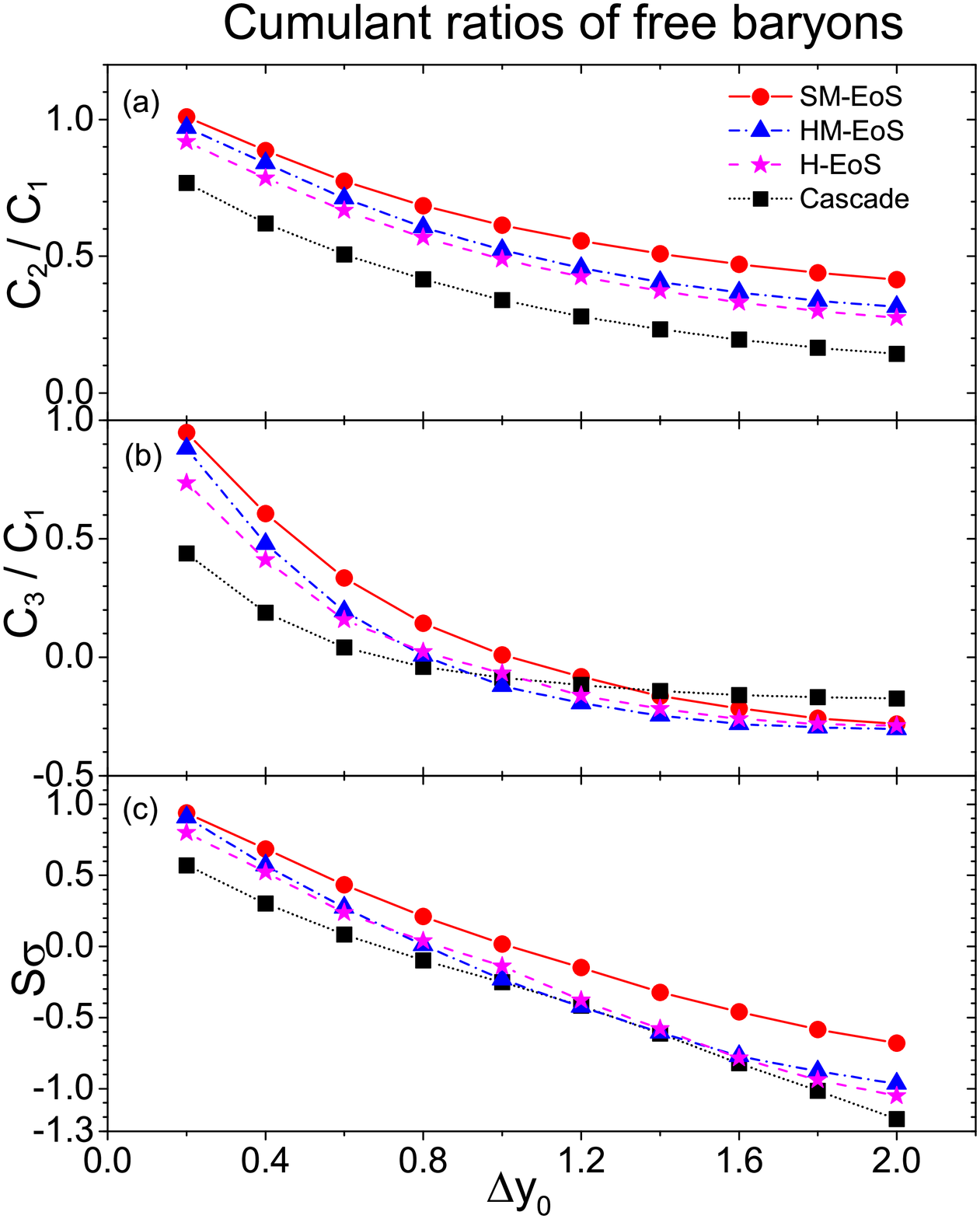}
\caption{\label{fig8} Rapidity dependence for the cumulant ratios ($C_2$/$C_1$, $C_3$/$C_1$, $S\sigma$) of free baryons, i.e. excluding all baryons that are in a cluster, produced in central Au+Au collisions at beam energy of 1.23 GeV/nucleon. Results have been calculated with SM, HM and H potentials as well as in the cascade mode (solid symbols). Error bars for all data shown, are smaller than the symbol sizes.}
\end{figure}

\begin{figure}[t]
\centering
\includegraphics[angle=0,width=0.5\textwidth]{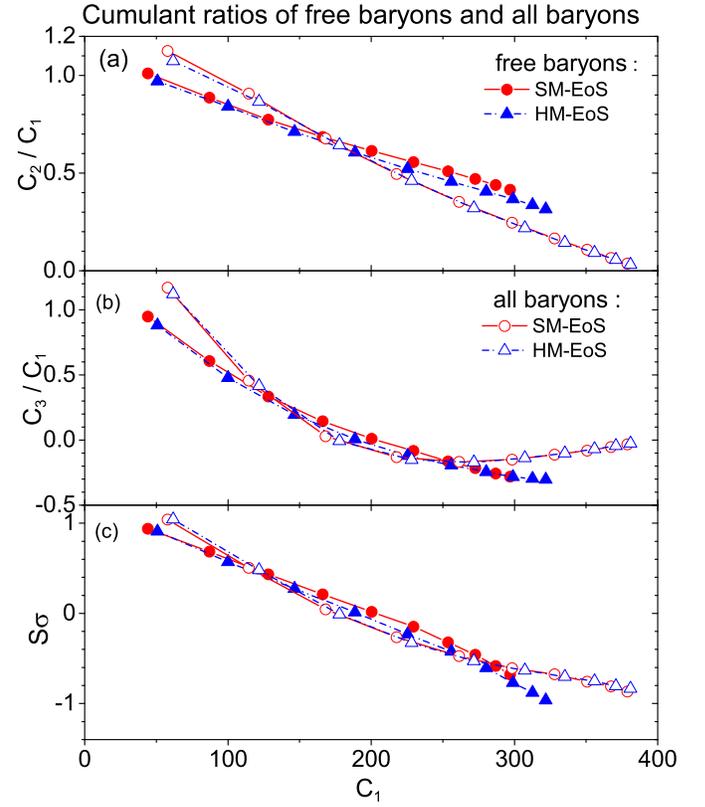}
\caption{\label{fig9} The cumulant ratios ($C_2/C_1$, $C_3/C_1$, $S\sigma$) of free baryons (solid symbols) and all baryons (open symbols) as a function of $C_1$. When plotted as function of $C_1$, the cumulant ratios for HM and SM parametrizations essentially agree with one another.}
\end{figure}

\begin{figure}[t]
\centering
\includegraphics[angle=0,width=0.5\textwidth]{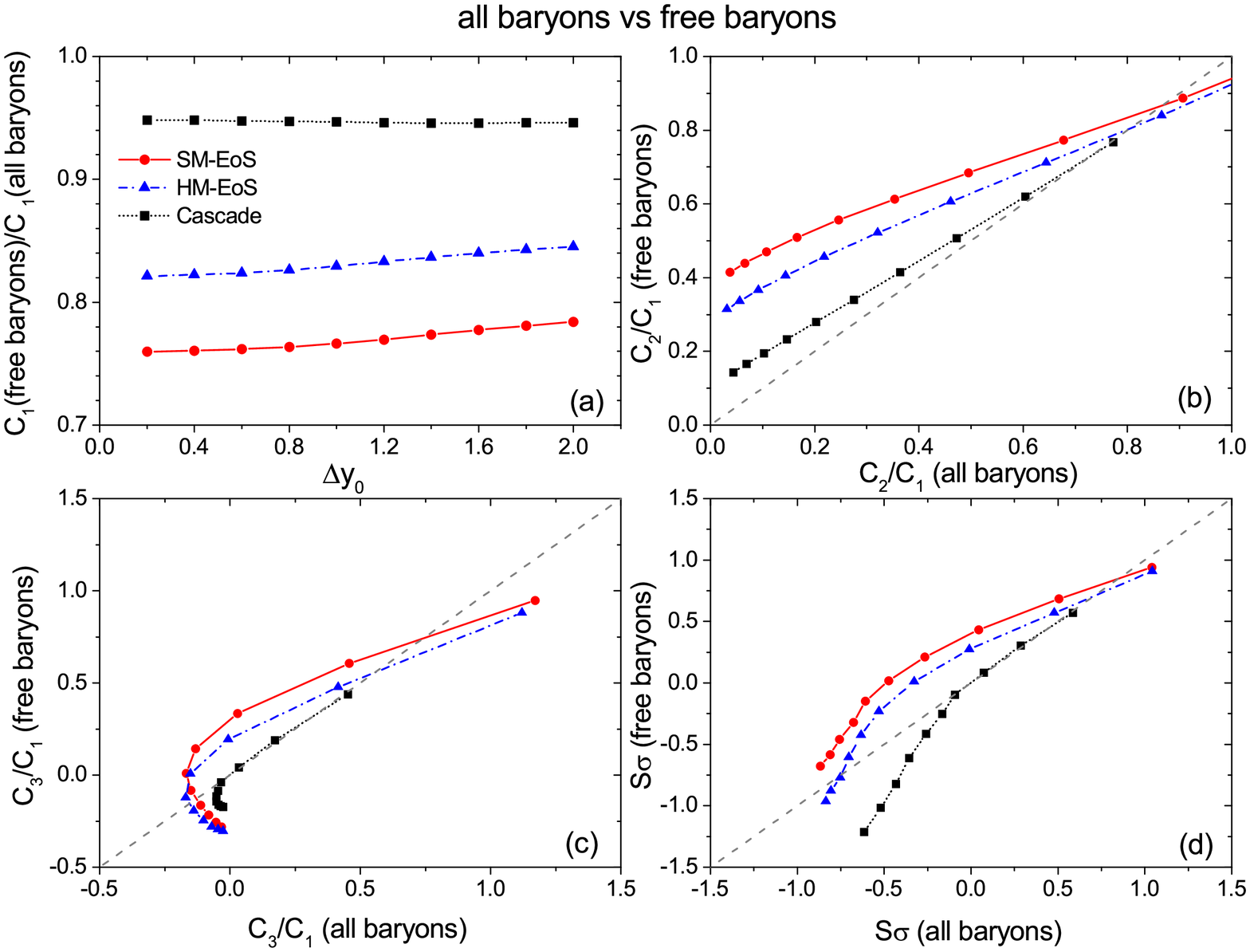}
\caption{\label{fig10} Comparison of the cumulant ratios for all baryons vs. free baryons. The diagonals are drawn to guide the eye. Mostly, an enhancement of the cumulant ratio with respect to all baryons is observed.}
\end{figure}

At the beam energy of 1.23 GeV$/$nucleon, multifragmentation is one of the main features and a large number of baryons belongs to fragments. In a previous study it was claimed that the formation of nuclear clusters can have a significant impact on the measured cumulant ratios \cite{Feckova:2015qza}.
However, the previous study was rather simplified and effects of conservation laws were neglected.
Thus it is important to study the cumulant ratios for free baryons (baryons that do not form a cluster) within a microscopic model as UrQMD. The results of our study are shown in Figures~\ref{fig7} and \ref{fig8}.
It can be seen that the magnitude of the cumulant ratios for free baryons is also enhanced by the mean field potential, as compared to the cascade simulation, similar to what was shown in Fig.\ref{fig2}.
A stronger attractive potential will thus yield more clusters.
The mean value ($C_1$) obtained from the cascade mode is the largest one and the one obtained with SM is the smallest one. On the other hand, the higher cumulants obtained from the cascade mode are the smallest, which is similar to the result for all baryons.\\

The general trends of the cumulant and cumulant ratios for free baryons are analogous to that for all baryons, but the cumulant ratios will no longer approach the binomial limit for large rapidity acceptances, since the free baryon number is no longer conserved.
On the other hand a clear difference between the calculation with the soft and hard momentum dependent potentials appears. Thus, the momentum dependence likely leads to a difference in cluster formation.

To understand this difference we show in Fig.~\ref{fig9}, the cumulant ratios for free baryons and all baryons as a function of $C_1$, nearly the same as in Figs.\ref{fig3} and \ref{fig8}, but the $\Delta y_0$ on the x-axis is replaced with the mean number ($C_1$).
We find that the difference in the HM and SM results are mainly due to the difference in $C_1$ which is caused by a different clusterization with various mean field potentials.

The difference in the cluster formation for the HM and SM potentials is shown in Fig.~\ref{fig10}. There also the results on the cumulant ratios for all baryons are directly compared to the ratios for free cumulants. If the results would lie on the diagonals, no effect of the clustering would be observed. Points that lie below the diagonals indicate a suppression of the cumulant ratio with respect to all baryons, while points that lie above the diagonal show an enhancement with respect to all baryons. One observes large deviations from the diagonals for all potential models. The ratio between the free baryons and all baryons in the case of SM show the largest deviations from the diagonals as there are 5 times more clusters created than in the cascade simulation.

Consequently, the cumulant ratios depend mainly on $C_1$.
The $C_1$ dependence also implies that the cumulant ratios for free baryons are less sensitive to the equation of state (EoS).
For small $C_1$ values (small $\Delta y_0$), the magnitude of the cumulant ratios for free baryons is smaller than that for all baryons, this qualitatively agrees with the result presented in Ref.~\cite{Feckova:2015qza}, where a strong reduction of cumulant ratio at midrapidity in the presence of deuteron formation was shown. However, for large values of $C_1$ the effect is reversed and the cluster formation leads to an increase in the cumulant ratio. This effect can be understood as a relaxation of the strict baryon number conservation for the free baryons which becomes more relevant for larger $\Delta y_0$.\\

These results manifests that the clusterization effect also plays an important role on the cumulant ratios of free baryons distributions.

\subsection{Results for free protons}

\begin{figure}[t]
\includegraphics[width=0.5\textwidth]{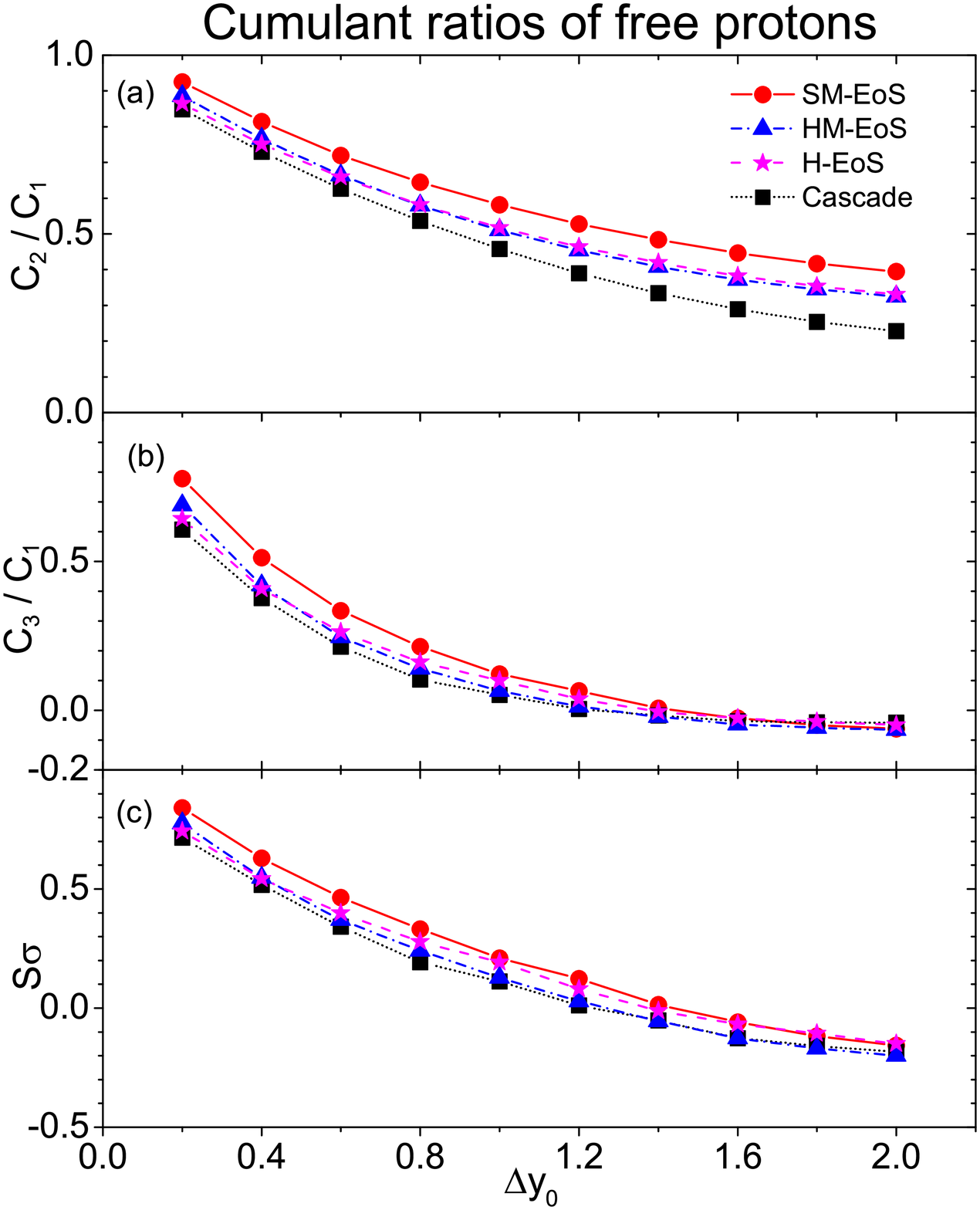}
\caption{\label{fig12} Rapidity dependence for the cumulant ratios of free protons (protons which are not bound in clusters) produced in central Au+Au collisions at beam energy of 1.23 GeV/nucleon. Results have been calculated with SM, HM and H potentials as well as in the cascade mode. Again, the differences between the cumulants for the different model setups is strongly decreased due to the almost random iso-spin distribution.  }
\end{figure}

In the CBM and HADES experiments, the fluctuation of net-proton number is used as the proxy observable for net-baryon number as they cannot measure neutrons. Thus, it is necessary to present the cumulant ratios for free protons, as shown in Fig.~\ref{fig12}. The cumulant ratios of free protons would be affected by the isospin randomization (e.g., neutron (proton) can be converted to proton (neutron) through the inelastic nucleon-nucleon collision) and the clusterization (i.e., a large fraction of protons is clustered in fragments).\\

In Ref~\cite{cover proton number to baryon number}, a set of formulas have been derived to convert the measured net-proton cumulants to the net-baryon cumulants by taking the effects of isospin exchange. In HICs, based on the assumption that the nucleons tend to completely forget their initial isospin, these formulas are expected to hold for $\sqrt{s_{NN}} > $10 GeV.
The assumption is likely not true anymore at intermediate energies due to the fact that the collision is not violent enough for the nucleons to completely lose the information on their initial isospin, i.e., $\langle N_{p} \rangle \ne \frac{1}{2}\langle N_{B} \rangle$.\\

In the UrQMD simulations, the cumulant ratios for free protons decrease monotonically with an increasing size of the rapidity window, which is similar to the bahaviour of the free baryons. However, the differences in the cumulant ratios for free protons between the different potentials becomes much smaller compared to that for free baryons or all baryons. The enhancements contributed from the mean field potentials still can be observed. Though, no cumulant ratios exceeds the value of 1.

\section{Summary}

The cumulant ratios for all baryons, free baryons, and free protons in central Au+Au collisions at a beam energy of 1.23 GeV$/$nucleon are investigated in the UrQMD and JAM models.
Calculations with the soft and hard momentum dependent (SM,HM) nuclear potential, and the hard potential without momentum dependence (H), are compared to each other and to the calculation without any mean field potential.

The cumulant ratios depend strongly on the reaction time: For early times, i.e. before 15 fm$/$c, the cumulant ratios obtained from HM, SM and cascade mode lie very close to each other in the momentum space.
Their magnitudes are all smaller than the Binomial baseline.
During the subsequent expansion stage, after 15 fm$/$c, the mean field potentials enhance the magnitude of the cumulant ratios.
This is predicted by both the UrQMD model and the JAM model.
A strong enhancement of the magnitude of the cumulant ratios of all baryons and of free baryons is clearly predicted for small rapidity acceptance when the mean field potentials are taken into account.
The enhancements are strongly reduced for free protons.
The cumulant ratios of all baryons are less sensitive to the density-dependent component of the nuclear potential but more sensitive to its momentum-dependent component.
The results of free baryons vs. those of all baryons show suppressed cumulant ratios for free baryons only at small rapidity windows.
Here the effects of the baryon number conservation are less important.
For large rapidity windows, clustering actually decreases the effects of the conservation laws.
Therefore, the cumulants increase.
Both the mean field, in particular its momentum dependent potential and the clustering play important roles in the cumulant ratios at intermediate energy heavy-ion collisions.

%----------------------------------------------------------------------------------------
%	CHAPTER 5
%----------------------------------------------------------------------------------------

\section{Acknowledgments}
The authors acknowledge the support of the computing server C3S2 at the Huzhou University. This work is supported in part by the National Natural
Science Foundation of China (Nos. 11875125, 11505057, 11747312, and 11375062), the Chinese Scholarship Council through the DAAD-PPP project, and the Zhejiang Provincial Natural Science Foundation of China (No. LY18A050002).
Y. N. acknowledges the support from the Grants-in-Aid
for Scientific Research from JSPS (Grants No. JP17K05448).
Horst Stoecker acknowledges the support through the Judah M. Eisenberg Laureatus Chair at Goethe University, and the Walter Greiner Gesellschaft, Frankfurt.

%----------------------------------------------------------------------------------------

\end{document}